\documentclass[10pt]{article}
\usepackage{fullpage}
\usepackage{amsmath}
\usepackage{amssymb}
\usepackage{mathtools}
\usepackage{setspace}
\begin{document}

\title{Lie Groups and their applications to Particle Physics: A Tutorial for Undergraduate Physics Majors}
\author{Jiaqi Huang}

\maketitle

\begin{abstract}

Symmetry lies at the heart of today’s theoretical study of particle physics. Our manuscript is a tutorial introducing foundational mathematics for understanding physical symmetries. We start from basic group theory and representation theory. We then introduce Lie Groups and Lie Algebra and their properties. We next discuss with detail two important Lie Groups in physics Special Unitary and Lorentz Group, with an emphasis on their applications to particle physics. Finally, we introduce field theory and its version of Noether's Theorem. We believe that the materials cover here will prepare undergraduates for future studies in mathematical physics.

\end{abstract}

\pagebreak

\tableofcontents
\newpage

%--INTRODUCTION--%
\section{Introduction: Physics From Symmetry}
Modern theoretical physics is a subject about symmetry: physicists \cite{Georgi} believes that the many physical principles in our observable universe are the result of symmetry breaking from some "theory of everything" that unifies all physical quantities as the different manifestations of the same quantity. More practically speaking, the study of symmetry provides us with ways to relate to physical observable and explain conservation law according to Noether's Theorem. Many modern-day theoretical physics problems such as the Yang-Mills theory \cite{Brink}, and the Grand-Unification theory \cite{Cham} relies on a correct understanding of symmetry. Thus, understanding symmetry is crucial for prospectives of theoretical physics research.
\\
\\
Group theory turns out to be a very nice platform for studying physical symmetry. The transformation groups are often useful to describe a full set of symmetry transformation with the same physical meaning. For example, we will see soon how the symmetry of a unit square can be represented by a group of $\frac{\pi}{2}$ unit rotations. Besides, the transformation groups can be represented as operators in some Hilbert Space, further proven their usefulness in describe symmetry in quantum mechanics where observables are described by operators. Concretely, given that most physical symmetries are differentiable symmetries, our study of group theory focus on those group whose group elements can be parametrized by some differentiable manifolds, groups which we called the Lie Groups. Further, since physics lives in vector spaces, we can concentrate only on the representations of the Lie groups. What's more, the infinitesimal generators of the Lie Group form a commutator Algebra, these generators are useful for defining the operator for the observables. We will learn these concepts in the following chapters to go.
\\
\\
In summary, the topics covered by this manuscript are the following. We will first start with finite group theory and introduces representation theory and many important concepts we will use in later chapters with finite groups. We will then move on to introducing Lie Group and Lie Algebras, and their important properties in the third chapter. In the fourth and fifth chapters, we will discuss in detail two important Lie Groups in physics the Special Unitary Groups and the Lorentz Group focusing on different representations of the Groups and their physical meaning. Finally, we will revisit the concept the readers might have learned in classical mechanics: Noether's Theorem, by introducing the version of it when used to describe the Lagrangian density of fields. Our goal is to prepare undergraduate and perhaps even some upper-level high school students with the appropriate background in classical mechanics, quantum mechanics, and Linear Algebra for advanced physical studies such as Quantum Field Theory and Particle Physics.

\section{Representation Theory in Finite Groups}
% Talks about groups, representation, group actions. An example of a finite group the Z2 and parity. Introducing vector/pseudo vector, scalar/pseudo scalar particles and Spin symmetry. 
We revisit concepts in introductory group theory courses that are important for the understanding of Lie groups and Lie Algebra. Besides, we provide examples in particle physics that illustrate these concepts.
\subsection{Groups: how it represents symmetry}
\begin{figure}[h]
        \centering
        \includegraphics[height = 5.0cm]{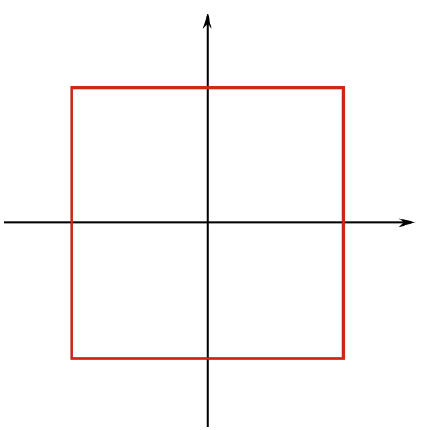}
        \caption{Illustration of Symmetry of a unit sqaure. When you rotate the unit square by any multiple of $\frac{\pi}{2}$, the shape of the unit square is preserved.}
        \label{square}
    \end{figure}
\noindent
\textbf{Definition 2.1.1} \textbf{Group} is a mathematical structure with the following properties. Let $f : G \times G \rightarrow G$ be a $\textbf{binary operation}$ 
\begin{enumerate}
    \item If $a,b \in G$, $c = f(a,b) \in G$ (Closure)
    \item For $a,b,c \in G, \ f(f(a,b),c) = f(a, f(b,c))$ (Associativity)
    \item There is an \textbf{Identity} element $e$, such that for all $a \in G, \ f(e,a) = f(a,e) = a$ (Identity)
    \item Every element $a \in G$ has an inverse $a^{-1} \in G$ such that $f(a, a^{-1}) = f(a^{-1}, a) = e$ (Inverse).
\end{enumerate}
To understand why a group is related to symmetry, let's start with a unit square. The unit square is a geometrical object that has a shape symmetry under any numbers of $\frac{\pi}{2}$ unit rotations. It turns out that there are only $\frac{\pi}{2}$ unit rotation transformations distinct $\pi$ unit rotation transformations out which can be denoted as $R = \{ R_{\frac{\pi}{2}}, R_{\pi}, R_{\frac{3\pi}{2}}, R_{0}\}$, since the effect of applying any number of $\frac{\pi}{2}$ unit rotation transformations can be achieved by only applying one of the 4 transformations described above, on any 2D Euclidean-Space geometrical objects. Let $f(a,b)$ be a binary operation which means first applying transformation a and then applying transformation b (eg. f($R_{\frac{\pi}{2}}$, $R_{\frac{3\pi}{2}}) = R_{0}$), one can immediately check that the set $R$ and the function f forms a group. Thus, the group of $\frac{\pi}{2}$ unit rotation transformations represent the shape symmetry of a unit square. When all groups element are transformations, we call the groups $\textbf{transformation groups}$, and it is these transformation groups which play an important role in explaining physical symmetries.

\subsection{Representations of Groups}
In Linear Algebra, we realize linear transformations as Matrices in different vector spaces of $\mathbb{R}^{n}$. It turns out that we can also represent symmetrical transformations in a group as linear operators in a vector space. 
\\
\\
\textbf{Definition 2.2.1} \textbf{Representation} of a group G is a mapping $g: G \rightarrow V$, for a vector space V, with the following properties:
\begin{enumerate}
    \item $g(e) = 1$, for 1 being the $\textbf{Identity operator}$ of the vector space.
    \item $g(a)g(b) = g(f(a,b))$, for $a,b \in G$ and f being the binary operation of G.
\end{enumerate}
Note that condition 1 implies that for each vector space, there exists only one representation for the group as there exists an unique identity operator in the vector space. As an example, consider the $\frac{\pi}{2}$ unit rotation transformations we described previously. Let the vector space be $\mathbb{R}^{2}$, then a representation of the group onto $\mathbb{R}^{2}$ with the standard basis can be the following map g:
\begin{align}
    & g(R_{0}) = g(e) = \begin{bmatrix}
    1 & 0\\
    0 & 1
    \end{bmatrix} & g(R_{90}) = \begin{bmatrix}
    0 & -1\\
    1 & 0
    \end{bmatrix}\\
    & g(R_{180}) = \begin{bmatrix}
    -1 & 0\\
    0 & -1
    \end{bmatrix} & g(R_{270}) = \begin{bmatrix}
    0 & 1\\
    -1 & 0
    \end{bmatrix}.
\end{align}
It will be an exercise for the readers to check that the above function $g$ is indeed a representation of the $\frac{\pi}{2}$ unit rotation transformations group. 
\\
\\
The representation of a group is not unique. It subjects the choice of basis of the vector space. However, when two representations differ only by the choice of basis, we call them \textbf{equivalent representations}. That is, let $D(g)$ and $D'(g)$ be two representations of the same vector space, then
\begin{equation}
    D'(g) = S^{-1} D(g) S,
\end{equation}
for some invertible transformation $S$. The following equation is also sometimes called the "change of basis operation" on D(g), and it turns out one can show that all representations map to the same vector space are equivalent, as one can always find a change of basis operation of $D(g)$ to get $D'(g)$.
\\
\\
However, when two representations map to two different vector spaces, they remain no longer equivalent. To illustrate this, consider the following representation $k$ of the $\frac{\pi}{2}$ unit rotation transformations group onto the vector space $\mathbb{R}^{4}$.
\begin{align}
    k(R_{0}) &= \begin{bmatrix}
    1 & 0 & 0 & 0\\
    0 & 1 & 0 & 0\\
    0 & 0 & 1 & 0\\
    0 & 0 & 0 & 1
    \end{bmatrix} & k(R_{90}) &= \begin{bmatrix}
    0 & 0 & 0 & 1\\
    1 & 0 & 0 & 0\\
    0 & 1 & 0 & 0\\
    0 & 0 & 1 & 0
    \end{bmatrix}\\
    k(R_{180}) &= \begin{bmatrix}
    0 & 0 & 1 & 0\\
    0 & 0 & 0 & 1\\
    1 & 0 & 0 & 0\\
    0 & 1 & 0 & 0
    \end{bmatrix} & k(R_{270}) &= \begin{bmatrix}
    0 & 1 & 0 & 0\\
    0 & 0 & 1 & 0\\
    0 & 0 & 0 & 1\\
    1 & 0 & 0 & 0
    \end{bmatrix}.
\end{align}
Again, one can check that k is indeed a representation of the $\frac{\pi}{2}$ unit rotation transformations group. The above representation is also called the \textbf{regular representation} of a group, as it is mapped to an Euclidean space with the same dimension as the order of the group. In turns out that there is a rule to generate these regular representations. Let b be the standard ortho-normal basis of the vector space $\mathbb{R}^{4}$. If we consider a bijective mapping B such that
\begin{align}
    &B(R_{0}) = b_{1} = \begin{bmatrix}
    1 \\
    0 \\
    0 \\
    0 
    \end{bmatrix} B(R_{90}) = b_{2} = \begin{bmatrix}
    0\\
    1\\
    0\\
    0\\
    \end{bmatrix}
    B(R_{180}) = b_{3} =\begin{bmatrix}
    0\\
    0\\
    1\\
    0\\
    \end{bmatrix} B(R_{270}) = b_{4} = \begin{bmatrix}
    0\\
    0\\
    0\\
    1\\
    \end{bmatrix},
\end{align}
then 
\begin{align}
    k(R)_{ij} = \langle b_{i}| k(R)| b_{j} \rangle,
\end{align}
where 
\begin{equation}
    k(R) b_{j} = B(f(R, B^{-1}(b_{j}))),
\end{equation}
for the binary function of the group f. The above scalar product in an Euclidean space is just the usual "dot product". Note that not all vector spaces have a scalar product, it is only those that are called \textbf{Hilbert Spaces} have a scalar product.

\subsection{Important Properties and Concepts}
We will introduce here some useful concepts which we will be using in the later chapters in the process of proving an important Theorem of the finite groups and their representations.
\\
\\
\textbf{Definition 2.3.1} \textbf{Algebra} V over a field is a vector space equipped with a billinear product $f : V \times V \rightarrow V$.
\\
\\
Note that the Algebra is defined over a field implies that the vector space is also equipped with addition, multiplication, and scalar multiplication from the field. 
\\
\\
The area of understanding groups by their representations and Algebra is called \textbf{representation theory}, which is also the area we will dive into when studying Lie Groups in the following chapters. One of the most useful aspects of representation theory lies in the following theorem. 
\\
\\
\textbf{Definition 2.3.1} Adjoint Operator. Let $V$ be a Hilbert Space, and consider $v, u \in V$, then the adjoint operator $A^{\dagger}$ of some operator $A$ is defined by
\begin{equation}
    \langle A^{\dagger}v, u \rangle =  \langle v, Au \rangle.
\end{equation}
In terms of matrices, the matrix of $A^{\dagger}$ is the conjugate transpose of the matrix of $A$. Note that if the Hilbert Space admits only real numbers as entries, then
\begin{equation}
    A^{\dagger} = A.
\end{equation}
\textbf{Definition 2.3.2} Unitary Operator of a Hilbert space $V$ is the operator with the following property:
\begin{equation}
    UU^{\dagger} = 1,
\end{equation}
where $U^{\dagger}$ is the adjoint operator of U.
\\
\\
\textbf{Theorem 2.3.1} Every representation of a finite group G is equivalent to a representation that maps all elements of G to a unitary operator in some Hilbert Space $V$.
\\
\textbf{Proof.} 
Suppose D(g) is a representation of a finite group G. Let
\begin{equation}
    S = \sum_{g \in G} D(g)^{\dagger} D(g).
\end{equation}
Note that since the operator $D(g)^{\dagger} D(g)$
is \textbf{Hermitian}, that is
\begin{equation}
    D(g)^{\dagger} D(g) = (D(g)^{\dagger} D(g))^{\dagger}.
\end{equation}
it automatically would have only real eigenvalues that diagnolize the vector space. It turns out that $D(g)^{\dagger} D(g)$ is also a \textbf{Positive Operator}, that is it has only real non-negative eigenvalues. This is true for arbitrarily chosen $D(g)$, and we will leave this as an exercise for the readers. It is then not hard to show that $S$ is also a $Positive Operator$ and we can diagonalize S as
\begin{equation}
    S = U^{-1} * d * U,
\end{equation}
where d is the diagonal matrix
\begin{equation}
    d = \begin{bmatrix}
    d_{1} & 0 & ...\\
    0 & d_{2} & ... \\
    ... & ... & ...,
    \end{bmatrix}
\end{equation}
where $d_{n} \geq 0$ are the eigenvalues. However, this doesn't guarantee that d is invertible as some $d_{n}$ may be 0. We now need to show that all $d_{n}$s must be non-zero. By contradiction, let $d_{i}$ be the element equal to zero, then there exists an eigenvector $v$ such that
\begin{equation}
    Sv = 0.
\end{equation}
But then
\begin{equation}
    v^{\dagger} S v = \sum_{g \in G} ||D(g) v||^{2} = 0,
\end{equation}
which means that every D(g) must maps to the zero matrix which is impossible since by the Definition 2.2.1
\begin{equation}
    D(e) = 1.
\end{equation}
Thus, S must be invertible. We then define 
\begin{equation}
    S^{\frac{1}{2}} = U^{-1} \begin{bmatrix}
    \sqrt{d_{1}} & 0 & ...\\
    0 & \sqrt{d_{2}} & ... \\
    ... & ... & ... 
    \end{bmatrix} U
\end{equation}
and let 
\begin{equation}
    D'(g) = S^{\frac{1}{2}} D(g) S^{\frac{1}{2}}{}^{-1}.
\end{equation}
Note that $S^{\frac{1}{2}}$ is automatically invertible given S is invertible. The only step left is to show that $D'(g)$ is indeed an unitary operator:
\begin{align}
    & D'(g)^{\dagger}D'(g) =  S^{\frac{1}{2}}{}^{-1} D(g)^{\dagger} S^{\frac{1}{2}} S^{\frac{1}{2}} D(g) S^{\frac{1}{2}}{}^{-1} \\
    & = S^{\frac{1}{2}}{}^{-1} D(g)^{\dagger} S D(g) S^{\frac{1}{2}}{}^{-1}\\
    & = S^{\frac{1}{2}}{}^{-1} D(g)^{\dagger} (\sum_{x \in G} D(x)^{\dagger} D(x)) D(g) S^{\frac{1}{2}}{}^{-1} \\
    & = S^{\frac{1}{2}}{}^{-1} \sum_{x \in G} D(gx)^{\dagger} D(gx) S^{\frac{1}{2}}{}^{-1} \\
    & = S^{\frac{1}{2}}{}^{-1} \sum_{h \in G} D(h)^{\dagger} D(h) S^{\frac{1}{2}}{}^{-1} \\
    & = S^{\frac{1}{2}}{}^{-1} S S^{\frac{1}{2}}{}^{-1} = 1.
\end{align}
% The above Theorem is extremely powerful, as it tells us that we can study ALL representations of a group by only studying their equivalent unitary representation. There are several convenience unitary operator can bring. First of all, unitarity automatically implies invertibility, and the inverse of an operator can be easily found as the adjoint operator of the operator. Secondly, the representation that maps every group element to an unitary operator is either \textbf{irreducible} or can be written as a \textbf{direct sum} of irreducible representations. 
% \\
% \\
% \textbf{Definition 2.3.3} An \textbf{Irreducible Representation} is representation with variant subspace. That is there exists an operator $D(g)$ and a subspace $V$ such that for some $v \in V$, $D(g) v \notin V$.
% \\
% \\
% \textbf{Definition 2.3.4} \textbf{Direct Sum} of representations is the mapping 
% \begin{equation}
%     D_{1} \opulus D_{2} = D_{3}, 
% \end{equation}
% where 
% \begin{equation}
%     Dim(D_{3}) = Dim(D_{1}) + Dim(D_{2}).
% \end{equation}
% In terms of vectors, the direct sum simply "concatenate" the vectors into a vector with the sum of the dimension of the vectors in the direct sum. 
% \\
% \\
% It is not immediately obvious why the second convenience of unitary representation is beneficial. In short, the convenience provides us the opportunity to only study the irreducible unitary representations, as any other unitary representation can be "formed" by the irreducibles, and thus equivalently to form any representation. 

\subsection{Example of Finite Group: Parity}
Are physical quantities the same in and out of the mirror the same? It turns out that some quantities are changed by a mirror reflection when others are not. For example, velocity changes its direction by 180 degrees on the other side of the mirror, while angular momentum remained unchanged by the mirror. Now let's define \textbf{helicity} as the quantity that determines whether a particle's spin angular momentum and velocity are in the same direction, that is, when helicity equals to 1, spin angular momentum and velocity are in the same direction, and -1 vice versa, then we know that a particle's helicity must change its sign under a mirror reflection. However, this is not true for a very interesting article that is the product of only weak interactions with the neutrinos. If we call -1 helicity particles as "left-handed" and + 1 as "right-handed", the discovery is that neutrinos are always "left-handed" on both sides of the mirror. If these results haven't struck you, I will tell you the implication: the mirror image of a neutrino simply doesn't exist! The unbelievable fact is also originally unbelievable to many great physicists such as Wolfgang Pauli who leaves his famous remark: "I can't believe God is left-handed." However, the statement is proven to be true after a series of verification from many different experiments. An assumption we made previously is that the neutrinos are massless, which may not be obvious at this point why this an important assumption. It turns out that because neutrinos are not massless and thus may not always travel at the speed of light \textbf{chirality} is a better way to define neutrino's "handedness". We will talk about chirality in the later chapter when we talk about Lorentz Groups.
\\
\\
Before we discuss with more details how such a claim was made, let's draw our attention back to Mathematics, and give a fancy name for mirror reflection - parity. Along with the identity operator, parity forms a trivial finite group as shown in the following multiplicative table:
\begin{table}[h]
\centering
\begin{tabular}{|l|l|l|}
\hline
\textbackslash{} & P & 1 \\ \hline
P                & 1 & P \\ \hline
1                & P & 1 \\ \hline
\end{tabular}
\end{table}
where each other element in the table is the product of the corresponding element in the first row and first column. It is clear that the inverse of P is itself as
\begin{equation}
    P^{2} = 1.
\end{equation}
One can also check given this fact that the eigenvalues of P are $\pm 1$. Thus, for any eigenfunction $\phi$ with eigenvalue 1, the function is unchanged under a mirror reflection and vice versa for eigenfunction with eigenvalue -1. The concept is the same as even and odd functions about a particular axis of reflection. To further distinguish scalar and vector function, physicists call even scalar functions and vector functions just as scalar and vector function, while calling odd scalar function and odd vector functions \textbf{pseudoscalar function} and \textbf{pseudovector function}. Physicists often classified particles into scalar/pseudoscalar and vector/pseudovector based on whether the wavefunction of the particles on a spin basis are scalar functions or vector functions. As a result, particles with non-zero spin are vector/pseudovector particles. Despite there are many spin 0 particles, there is only one known scalar particle so far: the famous Higgs Boson. In another word, all other spins 0 particles have odd wave functions on the spin basis. 
\\
\\
So what leads to the discovery of the broken mirror image of the neutrinos? An important fact to know is that neutrinos only participate in weak interactions, which describes most of the decays of the particles. The story all begins from the "tau-theta problem" proposed by the two Nobel Prize Winner: Yang and Lee \cite{Lee}. They discover that experimentally there are two particles $\tau^{+}$ and $\theta^{+}$ whose decay products are
\begin{equation}
    \tau^{+} \rightarrow \pi^{+} + \pi^{0} + \pi^{0},
\end{equation}
and
\begin{equation}
    \theta^{+} \rightarrow \pi^{+} + \pi^{0}.
\end{equation}
Let's treat each particle as some eigenstates of the Parity Operator and assume $\pi^{+}$ has parity +1 and $\pi^{0}$ has parity -1. What you will find is that $\tau^{+}$ is an eigenstate of the Parity Operator with eigenvalue -1 and $\theta^{+}$ with eigenvalue +1. However, the puzzle is that the two particles are identical except parity. Yang and Lee then proposed a theoretical explanation that the two particles must be the same while implying that parity is not conserved under the decays. Their theoretical proposal was later verified experimentally by Lee's colleague in Columbia, Wu \cite{Wu}, through studying the neutrinos produced by the decay of Cobalt-60 and found the shocking fact that neutrinos must always be left-handed.

\begin{figure}[h]
        \centering
        \includegraphics[height = 4.0cm]{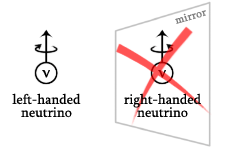}
        \caption{There's no right-handed neutrinos in nature.}
        \label{neutrinos}
    \end{figure}

\section{Lie Group and Lie Algebra}
% Talks about Lie Group, Lie Algebra, Simple group, and adjoint representations. Discuss physical meaning of roots and weights of the roots and weights, and raising and lowering operators.
The previous chapter focuses on finite groups, which are groups with finitely or countable many elements. However, since many physical symmetries are continuous symmetry, we must move on from finite groups to groups that can express these continuous symmetries. It turns out that there are groups whose elements are organized continuously by some continuous and smooth functions, which are exactly Lie Groups, the main focus of this paper.

\subsection{Lie Groups}
\textbf{Definition 3.1.1} Lie Group G = (S, f) is a group where S is a continuous set and f : $G \times G \rightarrow G$ is some smooth map. In another word, the Lie group forms a differentiable manifold.
\\
\\
Since we aim at explaining Lie group for students without a background in manifold theory, we will not explain concretely what a differentiable manifold is. Conceptually, one can consider differentiable manifold as shapes whose parametric equations are differentiable. For example, a unit square is not a differentiable manifold is but a unit circle is. As a result, to represent the symmetry of a unit circle, we need to introduce Lie groups. 
\\
\\
The symmetry of a unit circle is that its shape is invariant under unit rotations of \textbf{any degree}. The notion of any degree implies that there is continuously an infinite number of unit rotations one can perform such that the unit circle's shape is unchanged. To do so we need some continuous variable denoted $\theta$ to represent every possible unit rotations. Now we call these unit rotations $R$, and consider the representation of these unit rotations in $\mathbb{R}^{2}$ then,
\begin{equation}
    R(\theta) = \begin{bmatrix}
        \cos(\theta) & -\sin(\theta) \\
        \sin(\theta) & \cos(\theta).
    \end{bmatrix}
\end{equation}
Note that when $\theta = 0$, we have
\begin{equation}
    R(0) = 1.
\end{equation}
Now consider some neighborhood of the identity operator, we can perform a Taylor Expansion and write
\begin{equation}
    \label{32}
    R(d\theta) \approx 1 + R'(0)d\theta,
\end{equation}
where we keep only the first two term assuming $d\theta$ is small enough. Note that 
\begin{equation}
    R'(\theta) = \begin{bmatrix}
        -\sin(\theta) & -\cos(\theta) \\
        \cos(\theta) & -\sin(\theta).
    \end{bmatrix}
\end{equation}
So
\begin{equation}
    R'(0) = \begin{bmatrix}
        0 & -1 \\
        1 & 0
    \end{bmatrix}.
\end{equation}
The above matrix $R'(0)$ yields only imaginary eigenvalues. As physicists, we want our operators to be Hermitian, since in physics want to make use of the eigenvalues to express conserved quantities in the symmetry. Thus, we define the \textbf{generator} of the unit rotation Lie group as following:
\begin{equation}
    \label{35}
    X = -iR'(0).
\end{equation}
There is a actually a very clear reason why we called $X$ the generator of the Lie group. Since we can write
\begin{equation}
    R(d\theta) \approx 1 + id\theta X,
\end{equation}
for any infinitesimal element near the identity operator according Eq. \ref{32} and Eq. \ref{35}, and also since we know that the unit rotations in $\mathbb{R}^{2}$ forms a group, any numbers of matrix multiplication of $D(d\theta)$ takes away from the identity to any other group elements. Thus, infinite numbers of applications should generate all unit rotations in $\mathbb{R}^{2}$, and we can write
\begin{equation}
    R(\theta) = \lim_{k \rightarrow \infty} (1 + id\theta X)^{k} = \lim_{k \rightarrow \infty} (1 + i\theta X/k)^{k} = e^{iX\theta}.
\end{equation}
If you have a Mathematica or any mathematical computation programming language with you you can check that $e^{iX\theta}$ indeed gives us $R(\theta)$ as expected. 
\\
\\
In general, we can define the generators as the following.
\\
\\
\textbf{Definition 3.1.2} Generators of Lie Group. Let $\alpha_{j}$ be a set of parameters that parametrize the Lie group, and let D be a representation of the Lie group such that
\begin{equation}
    D(0) = 1.
\end{equation}
Then the generators of the Lie Group are
\begin{equation}
    X_{j} = - i \frac{\partial D}{\partial \alpha_{j}} (0). 
\end{equation}
In such a case,
\begin{equation}
    D(\alpha) = e^{iX_{j}\alpha^{j}}
\end{equation}
Note that the index notation above implies that when there are more than one generator, we have the general generator as a linear combination generators. If you haven't learnt how to deal with index notation and Einstein Summation, please refer to \cite{Ben} for some tutorials of it. 
\\
\\
You may wonder why do we want to study Lie group through their generators. The truth is that the generators form an Algebra under matrix commutation. You will see appreciate this fact as we move on and on further. 

\begin{figure}[h]
        \centering
        \includegraphics[height = 6.0cm]{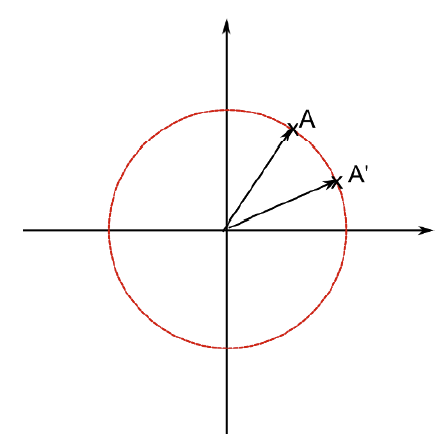}
        \caption{Illustration of the symmetry of a unit circle. The shape of circle is invariant under rotation of any angle.}
        \label{neutrinos}
    \end{figure}

\subsection{Lie Algebra: Lie Bracket and Jacobi Identity}
Now we will show that the generators indeed form an Algebra. First of all, it is observed that the generators must all be operators in the same vector space the representation of the Lie Group maps to, so they must also form an operator vector space just like the group elements themselves. The only thing we need to check is that there exists a bilinear product in the vector space of the generators. The fact is that there is such a bilinear product called matrix commutator, and we will check this now.
\\
\\
Suppose we start with two group elements $e^{i\alpha_{a}X_{a}}$ and $e^{i\beta_{b}X_{b}}$, where $\alpha_{a}$ and $\beta_{b}$ are just two different set of parameters for the Lie group, then by Group Closure,
\begin{equation}
    e^{i\alpha^{a}X_{a}}e^{i\beta^{b}X_{b}} = e^{i\delta^{c}X_{c}},
\end{equation}
for some new element $e^{i\delta_{c}X_{c}}$. We can then write
\begin{equation}
    i\delta^{c}X_{c} = \ln ( e^{i\alpha^{a}X_{a}}e^{i\beta^{b}X_{b}}) = \ln (1 +  e^{i\alpha^{a}X_{a}}e^{i\beta^{b}X_{b}} - 1).
\end{equation}
The benefit of appending and subtracting a 1 is that we can then use the Taylor Expansion of $\ln(1 + K)$ for $K = e^{i\alpha^{a}X_{a}}e^{i\beta^{b}X_{b}} - 1$ to approximate new linear combination of generators $i\delta^{c}X_{c}$. We first approximate K by a binomial expansion
\begin{align}
    &K = e^{i\alpha^{a}X_{a}}e^{i\beta^{b}X_{b}} - 1 \\
    &= (1 + i\alpha^{a}X_{a} - \frac{1}{2}(\alpha_{a}X_{a})^{2} + ...)(1 + i\beta^{b}X_{b} - \frac{1}{2}(\beta_{b}X_{b})^{2} + ...) - 1 \\
    & = i\alpha^{a}X_{a} + i\beta^{b}X_{b} - \alpha^{a}X_{a}\beta^{b}X_{b} - \frac{1}{2}(\alpha^{a}X_{a})^{2} - \frac{1}{2}(\beta^{b}X_{b})^{2} + ...
\end{align}
Thus,
\begin{align}
    & i\delta^{c}X_{c} = \ln(1 + K)\\
    & = K - \frac{1}{2}K^{2} + ... \\
    &  = i\alpha^{a}X_{a} + i\beta^{b}X_{b} - \alpha^{a}X_{a}\beta^{b}X_{b} - \frac{1}{2}((\alpha^{a}X_{a})^{2} - (\beta^{b}X_{b})^{2}) + \frac{1}{2}(\alpha^{a}X_{a} + \beta^{b}X_{b})^{2} + ...
\end{align}
Note that the term:
\begin{equation}
    \frac{1}{2}((\alpha^{a}X_{a})^{2} - (\beta^{b}X_{b})^{2}) + \frac{1}{2}(\alpha^{a}X_{a} + \beta^{b}X_{b})^{2} = \frac{1}{2}[\alpha^{a}X_{a}, \beta^{b}X_{b}],
\end{equation}
by manipulating the equation above, we can write
\begin{equation}
    [\alpha^{a}X_{a}, \beta^{b}X_{b}] = -2i((\delta^{c} - \alpha^{c} - \beta^{c})X_{c} + ....),
\end{equation}
Note that we can rewrite the indexing variable of the index notation because they are in different terms. An important observation to make here is that, no matter what the infinite sum in the right-handside will be, it must be some linear combination of the generators multiply by $i$. Let's called linear combination $\gamma^{c}X_{c}$, we have that
\begin{equation}
    [\alpha^{a}X_{a}, \beta^{b}X_{b}] = i\gamma^{c}X_{c}.
\end{equation}
And now let's define 
\begin{equation}
    \gamma^{c} = \alpha^{a}\beta^{b}f_{ab}^{c},
\end{equation}
we have
\begin{equation}
    \label{53}
    [X_{a}, X_{b}] = if_{ab}^{c}X_{c}.
\end{equation}
The $f_{ab}^{c}$ are called the \textbf{structral constants} of the Lie Algebra. Even if we call them constants, $f_{ab}^{c}$ is really a three-tensor of constants. 
\\
\\
The implication of Eq. \ref{53} is that for arbitrarily given $X_{a}, X_{b}, X_{c}$, we can map from any two of the three generators to another one using matrix commutation. Thus, matrix commutator does form a bilinear product of the vector space of generators. The matrix commutators are often given a fancy name called \textbf{Lie Bracket} in order to distinguish with another possible billinear product for the generators \textbf{Poisson Bracket} which is 
\begin{equation}
    \{X_{a}, X_{b}\} = \frac{\partial X_{a}}{\partial\alpha_{a}} \frac{\partial X_{b}}{\partial\alpha_{b}} - \frac{\partial X_{a}}{\partial\alpha_{b}} \frac{\partial X_{b}}{\partial\alpha_{a}} = \frac{\partial X_{a}}{\partial\alpha_{a}} \frac{\partial X_{b}}{\partial\alpha_{b}},
\end{equation}
given the specific way we create these generators. In physics, Poisson Bracket is often used when we are working with classical Hamilton's equations. When the Poisson Bracket is replaced by Lie Bracket, it automatically implies the Hamiltonian or the Lagrangian is mapped to some Hilbert Space where the matrix algebra works and the full Hamiltonian Equation of Motion reduces to the \textbf{Heisenberg picture of Quantum Mechanics}. That is given some observables $A$ (energy, momentum etc) and the Hamiltonian $H$ 
\begin{equation}
    \frac{d}{dt} A(t) = \frac{i}{\hbar}[A, H] + (\frac{\partial A}{\partial t})_{H},
\end{equation}
Just a recap in case you forget, in Classical Mechanics, the classical Hamilton's Equation is
\begin{equation}
    \frac{d}{dt} f(p,q,t) = \frac{i}{\hbar}{f(p,q,t), H(p,q,t)} + (\frac{\partial f}{\partial t})_{H},
\end{equation}
for some continuous function $f$ mapping the systems generalized coordinate, momentum, and time to some real-value physical quantity. Note that in Hamilton's equation, $H$, the hamiltonian, is a three parameter continuous function, while $H$ in the Heisenberg picture is an operator. This indicates that the quantum mechanical Hamiltonian acts on continuous functions rather than being a continuous function itself.
\\
\\
For the sake of simplicity, we will consistently assume now that the Lie Bracket is the bracket we use for the Lie Algebras.
\\
\\
The important properties of the Lie Bracket are the following:
\begin{enumerate}
    \item \begin{equation}
        [\alpha^{a}X_{a}, \alpha^{a}X_{a}] = 0
    \end{equation}
    \item \begin{equation}
        [\alpha^{a}X_{a} + \beta^{b}X_{b}, \gamma^{c}X_{c}] = [\alpha^{a}X_{a}, \gamma^{c}X_{c}] + [\alpha^{b}X_{b}, \gamma^{c}X_{c}]
    \end{equation}
    \item
    \begin{equation}
        [\alpha^{a}X_{a}, \beta^{b}X_{b}] = - [\beta^{b}X_{b}, \alpha^{a}X_{a}]
    \end{equation}
    \item 
    \begin{equation}
        [\alpha^{a}X_{a}, [\beta^{b}X_{b}, \gamma^{c}X_{c}]] + [\beta^{b}X_{b}, [\gamma^{c}X_{c}, \alpha^{a}X_{a}]] + [\gamma^{c}X_{c}, [\alpha^{a}X_{a}, \beta^{b}X_{b}]] = 0.
    \end{equation}
\end{enumerate}
The first three are rather easy to check to true given the definition of Lie Bracket. The last one which is not as obvious is often called \textbf{Jacobi Identity} of the Lie Bracket. Note that the way we write the three linear combinations of generators in the nested brackets must be \textbf{cyclic permutation} of one another, meaning that one permutation of the three linear combinations can only be obtained by cyclically rotating the position of the linear combinations of another permutation. Although the Jacobi Identity doesn't look too obvious it is indeed just another way of writing the product rule of the commutators
\begin{equation}
    [X_{a}, X_{b}X_{c}] = [X_{a}, X_{b}]X_{c} + X_{b}[X_{a}, X_{c}].
\end{equation}
It will leave as an exercise for the readers to prove Jacobi Identity from the product rule.
\\
\\
\textbf{Definition 3.2.1} Lie Algebra of a Lie Group is the Algebra formed by the group's generators. The bilinear product is usually Lie Bracket or Poisson Bracket.

\subsection{Example of Lie Group: Angular Momentum and $SO(3)$}
Previously, we talk about unit rotations in two-dimensional Euclidean Space look like. We will now examine how a three-dimensional unit rotation can be represented. Before that, we will formally introduce the unit rotation group.
\\
\\
\textbf{Definition 3.3.1} unit rotation Groups SO(n) are groups of all n dimensional unit unit rotations. 
\\
\\
Thus, the previous unit rotation in two-dimensional is the $SO(2)$ group. However, it is very important to understand that a three-dimensional representation of $SO(2)$ is different from $SO(3)$. Although being represented in a 3D Euclidean Space, the three-dimensional representation of $SO(2)$ must still lies in some two-dimensional subspace of the 3D Euclidean Space. For example, an $SO(2)$ group in $\mathbb{R}^{3}$ is
\begin{equation}
R(\theta) = 
\begin{bmatrix}
    1 & 0 & 0 \\
    0 & \cos\theta & -\sin\theta \\
    0 & \sin \theta & \cos\theta
\end{bmatrix},
\end{equation}
which is unit rotations along the axis of unit vector (1,0,0). However, this group does not represent all three dimensional unit rotations, as there can be infinite number of axes to be rotated on a 3D Euclidean Space. Since we know that all unit rotations axes can be spanned by coordinates in $x,y, z$ axes, we can consider the following basis of unit rotation along the $x, y, z$ axes as:
\begin{align}
R_{x}(\theta) &= 
\begin{bmatrix}
    1 & 0 & 0 \\
    0 & \cos\theta & -\sin\theta \\
    0 & \sin \theta & \cos\theta
\end{bmatrix}, &
R_{y}(\theta) &= 
\begin{bmatrix}
    \cos\theta & 0 & \sin\theta \\
    0 & 1 & 0 \\
    -\sin\theta & 0 & \cos\theta
\end{bmatrix}, & R_{z}(\theta) &=
\begin{bmatrix}
    \cos\theta & -\sin\theta & 0 \\
    \sin \theta & \cos\theta & 0\\
    0 & 0 & 1
\end{bmatrix}.
\end{align}
Note that this is only one possible basis. There can be some other basis that span all three-dimensional unit rotations as well. The Generators of the Lie Group given the three basis are therefore
\begin{align}
X_{x} &= 
\begin{bmatrix}
    0 & 0 & 0 \\
    0 & 0 & i \\
    0 & -i & 0
\end{bmatrix}, & X_{y} &= 
\begin{bmatrix}
    0 & 0 & -i \\
    0 & 0 & 0 \\
    i & 0 & 0
\end{bmatrix}, &
X_{z} &= 
\begin{bmatrix}
    0 & i & 0 \\
    -i & 0 & 0\\
    0 & 0 & 0
\end{bmatrix}.
\end{align}

The Eigenvalues of all of the above generators are $\{1, 0, -1\}$. As we described briefly that the eigenvalues are conserved charges of a symmetrical transformation, which in this case is unit rotation, the eigenvalues are thought to be some measurements of angular momentum. In fact, these are generators correspond to particle with spin angular momentum $\hbar$, which is usually also called spin 1 particles. To be clear, we define the angular momentum operators as
\begin{equation}
    J_{a} = \hbar X_{a},
\end{equation}
such that the eigenvalues are exactly $\{\hbar, 0, -\hbar\}$. The unit rotations are therefore 
\begin{equation}
    R(\alpha^{a}) = e^{i\frac{\alpha^{a}}{\hbar}J_{a}},
\end{equation}
where $\alpha^{a}$ is the vector with value for each of the three unit rotation angles. 
\\
\\
Finally, we check that the Generators of $SO(3)$ does form an Lie Algebra, with 
\begin{equation}
    [X_{i}, X_{j}] = 2i\epsilon_{ijk}X_{k},
\end{equation}
where $\epsilon_{ijk}$ is the \textbf{Levi-Civita Symbol}
\begin{equation}
    \epsilon_{ijk} = \begin{cases}
         +1 & \text{if } (i,j,k) \text{ is } (1,2,3), (2,3,1), \text{ or } (3,1,2), \\
         -1 & \text{if } (i,j,k) \text{ is } (3,2,1), (1,3,2), \text{ or } (2,1,3), \\
         0 & \text{if } i = j, \text{ or } j = k, \text{ or } k = i.
    \end{cases}
\end{equation}
It is striking but true that, though there doesn't exist a two-dimensional representation of a three-dimensional rotator, there exists a two-dimensional representation of some group whose Algebra is the same as $SO(3)$. The group is called $SU(2)$, which describes objects in a completely different functional space - the \textbf{Complex Space}. As you can imagine, things are a lot different if our matrices representation of the group elements can accept complex numbers. We will discuss more the reason why we can "simulate" three-dimensional unit rotations with a 2D Complex Space when we formally introduce the \textbf{Special Unitary Groups}. The intuition behind this is the following example. Consider $SO(2)$, if we include complex numbers, the cosine part can be the real part of the complex number, and the sine part can be the imaginary part. We can then somehow "simulate" the two-dimensional unit rotations as some kind of one-dimensional complex number.

\newpage

\section{Special Unitary Groups}
% $SU(2)$ and its representations. Example: Isospin Symmetry. SU(3) and its representations. Example: Hypercharge Symmetry and Gell-Mann Okobu Formula. SU(5) and unification theory.
We talk about in the previous chapter that there exists some two-dimensional representation of the $SO(3)$ Algebra. It may sound a bit weird when we talk about a two-dimensional representation of the 3D unit rotations. In the following chapter, we will unveil the reason why by introducing the Special Unitary Group which is a transformation group in complex space rather than in Euclidean Space. 
\subsection{$SU(2)$ is the Double Cover of $SO(3)$}
To understand the reason why can have a two-dimensional of the Algebra of $SO(3)$ in complex space, we can first consider how could we represent $SO(2)$ using complex numbers. As you could probably observe, it isn't hard to construct such a complex number if we align the real part and imaginary part of a complex number with the matrix
\begin{align}
1 &= 
    \begin{bmatrix}
        1 & 0 \\
        0 & 1
    \end{bmatrix} & i & = 
    \begin{bmatrix}
        0 & -1 \\
        1 & 0
    \end{bmatrix},
\end{align}
and the imaginary part with the matrix
Then the entire $SO(2)$ can be expressed by the following complex number equation
\begin{equation}
    R(\theta) = \cos(\theta) + \sin(\theta).
\end{equation}
The above complex number equation also forms a different Lie Group which is $U(1)$, the unitary group of one-dimensional complex numbers.
\\
\\
\textbf{Definition 3.1.1} Unitary Group U(n) is the group of n dimensional complex unitary transformations.
\\
\\
In our case, it is not hard to check that the one-dimensional $R(\theta)$ is unitary given that the adjoint of $R(\theta)$ is just its complex conjugate. An important observation is that the real number 1 is the generator of $U(1)$ since by Euler's Formula:
\begin{equation}
    e^{i*1\theta} = cos(\theta) + i sin(\theta).
\end{equation}
The reason why we can easily build a complex version of the Algebra of $SO(2)$ is that the entire symmetry group relies on only two degrees of freedom so we can effectively use a single complex number to capture the degree of freedom. However, two degrees of freedom is not enough for $SO(3)$. As a result, an educated guess would be to look at higher dimensional unitary operations. The next dimension is 2 which leads us to the group $U(2)$. Now we claim that only those unitary operators with determinant 1 will be sufficient to model the Algebra of $SO(3)$ given that $SO(3)$ has determinant 1 as well. Without loss of generality, last call these determinant 1 unitary operator "Special Unitary" operators which forms a symmetry group denoted as $SU(2)$. Consider the usual two-dimensional representation of $U(2)$, because of the property of unitary operators and the requirement that the determinant is 1, one can show that matrix element of the special unitary operators are
\begin{equation}
    SU(2) = \{ \begin{bmatrix}
        \alpha & -\beta^{*}\\
        \beta & \alpha^{*}
    \end{bmatrix} : \alpha, \beta \in \mathbb{C}, |\alpha|^{2} + |\beta|^{2} = 1 \}.
\end{equation}
The implication of the matrix elements of $SU(2)$ in the usual two dimensional representation is that $SU(2)$ gives a maximum of four degrees of freedom, with two degree of freedom for each complex number. We then need four basis operators to write all operators in $SU(2)$. Consider the following four matrices as our choice of basis
\begin{align}
    1 & = \begin{bmatrix}
    1 & 0\\
    0 & 1 
    \end{bmatrix} & I & = \begin{bmatrix}
    0 & i \\
    i & 0
    \end{bmatrix}\\ J & = \begin{bmatrix}
    0 & 1\\
    -1 & 0 
    \end{bmatrix} & K & = \begin{bmatrix}
    i & 0\\
    0 & -i 
    \end{bmatrix}  .
\end{align}
Then we observe that
\begin{equation}
    J^{2} = I^{2} = K^{2} = -1,
\end{equation}
and every other element $m$ of $SU(2)$ can be written as 
\begin{equation}
    m = a + bI + cJ + dK,
\end{equation}
for some real coefficients $a,b,c,d$, if these real coefficients form a convex combination. The way we construct these indexes automatically gives us some way of writing the operators in the form some "complex number with three imaginary axes". We call these new "complex number" \textbf{Quaternion}. 
\\
\\
At this point, it still isn't obvious what a mapping that simulates three-dimensional unit rotation would look like using $SU(2)$. There are two reasons. First of all, we still didn't express the group as an operator of some continuous variables. Second, a Quaternion or $SU(2)$ has a total of 4 degrees of freedom while $SO(3)$ has three for each different angle. The solution to the first problem can be solved by parametrizing the coefficients of each axis of the Quaternion and the solution to the second problem can be solved by fixing the coefficient in front of one of the axes.
\\
\\
In particular, consider the following Quaternions:
\begin{equation}
    t(\theta) = \cos\theta + \sin\theta u,
\end{equation}
where
\begin{equation}
    u = x_{q}I + y_{q}J + z_{q}K,
\end{equation}
for $x^{2} + y^{2} +z^{2} = 1$. It is not hard to check that the above expression does gives a Quaternions and does express three degrees of freedom given that we fix the coefficient in front of the real part as $\cos\theta$. Now let's express align the x,y,z axes of an Euclidean Space vector with I, J, and K, then t looks like some kind of unit rotation about an axis of [$x_{q}$, $y_{q}$, $z_{q}$]. We can then express our $SO(3)$ basis operators $R_{x}(\theta)$, $R_{y}(\theta)$, and $R_{z}(\theta)$ as correspondingly
\begin{equation}
    t_{x}(\theta) = cos\theta + \sin\theta I
\end{equation}
\begin{equation}
    t_{y}(\theta) = cos\theta + \sin\theta J
\end{equation}
\begin{equation}
    t_{z}(\theta) = cos\theta + \sin\theta I
\end{equation}
Suppose we construct a mapping for any Euclidean vector in 3D to the Quaternion:
\begin{equation}
    \begin{bmatrix}
        x\\
        y\\
        z
    \end{bmatrix} = xI + yJ +zK \equiv m,
\end{equation}
one can check that
\begin{equation}
    m' = t_{x}(\theta)m
\end{equation}
doesn't gives you the correct unit rotation along the axis if we map the resulting Quaternion back to the 3D Euclidean Space by the inverse of the above mapping. Indeed the correct unit rotations is
\begin{equation}
    m' = t_{x}(\theta)mt_{x}^{-1} = t_{x}(\theta)mt_{x}^{\dagger}(\theta),
\end{equation}
given that $R_{x}(\theta)$ is unitary operator. In general, for any 3D unit rotation $R_{\theta}$, we can write the following Quaternion Operation
\begin{equation}
    m' = t(\theta)mt^{\dagger}(\theta),
\end{equation}
to simulate the unit rotation operation. In the form of $2 \times 2$ matrices, this equation becomes
\begin{equation}
    \begin{bmatrix}
        x'i & -y' + z'i \\
        y' + z'i & -x'i
    \end{bmatrix} = \begin{bmatrix}
        \cos\theta + \sin\theta x_{p}i & \sin\theta(-y_{p} + z_{p}i) \\
        \sin\theta (y_{p} + z_{p}i) & \cos\theta - \sin\theta x_{p}
    \end{bmatrix} \begin{bmatrix}
        xi & -y + zi \\
        y + zi & -xi
    \end{bmatrix} \begin{bmatrix}
        \cos\theta - \sin\theta x_{p}i & \sin\theta(y_{p} - z_{p}i) \\
        \sin\theta (-y_{p} - z_{p}i) & \cos\theta + \sin\theta x_{p}
    \end{bmatrix}.
\end{equation}
It is quite important to note that the mapping from $R(\theta)$ to t is not one to one: we can write $t(\theta)$ as $t^{\dagger}(\theta)$ to result in the same simulation. This makes intuitive sense since $SU(2)$ and $SO(3)$ are not exactly the same group given that they didn't have the same maximum degree of freedom. Because of we can map two t to one $R(\theta)$, it makes sense that $SU(2)$ is the \textbf{Double Cover} of $SO(3)$. 
\\
\\
\textbf{Definition 4.3.2} Double Cover. Let G and H be two homomorphic groups. G is the double cover of H if exactly two group elements of G maps to one element of H under the group Homomorphism.
\\
\\
Through the above simulation of $SO(3)$ in $SU(2)$, we have proven the Double Cover relation between the two groups.
\\
\\
Perhaps it is more striking than if we consider a convex combination of $t_{x}$, $t_{y}$, and $t_{z}$, we indeed span all elements of the entire $SO(3)$. How come we previously use four matrices to span the entire group but now we can use three? The reason lies in the flexibility complex numbers give. Recall the definition of Special Unitary Groups in Definition 3.1.1, we observe that the matrices can result in any degree of freedom between 4 and 2, given that we can write both real and imaginary part of a complex number as a single one-dimensional number. The same idea applies here, as if we examine more closely, each one of $t_{x}$, $t_{y}$, and $t_{z}$ two of the Quaternions, thus makes sense to span the entire group as through some convex combination all possibility of each axis of the Quaternion is represented. It is important to clarify here, though this is true, $SO(3)$ and $SU(2)$ are still not isomorphic given that what we can manipulate in $SO(3)$ are analogously the three imaginary parts of the Quaternion as we use them to represent the 3D Euclidean vectors, so any change unit rotation of $SO(3)$ must still map to only Quaternions with $\cos\theta$ in the real part.
\\
\\
We now can examine the generators of the above two dimensional representation of $SU(2)$:
\begin{equation}
    \sigma_{x} = -i\frac{dt_{x}}{d\theta}(0) = -i\begin{bmatrix}
     -\sin(0) & i\cos(0)\\
     i\cos(0) & -\sin(0)
    \end{bmatrix} = \begin{bmatrix}
     0 & 1\\
     1 & 0
    \end{bmatrix}
\end{equation}
\begin{equation}
    \sigma_{y} = -i\frac{dt_{y}}{d\theta}(0) = -i\begin{bmatrix}
     -\sin(0) & \cos(0)\\
     -\cos(0) & -\sin(0)
    \end{bmatrix} = \begin{bmatrix}
     0 & -i\\
     i & 0
    \end{bmatrix}
\end{equation}
\begin{equation}
    \sigma_{z} = -i\frac{dt_{y}}{d\theta}(0) = -i\begin{bmatrix}
     -\sin(0) + i\cos(0) & 0\\
     0 & -\sin(0) - i\cos(0
    \end{bmatrix} = \begin{bmatrix}
     1 & 0\\
     0 & -1
    \end{bmatrix}
\end{equation}
These generators may look familiar to a lot of you. These are the \textbf{Pauli Matrices} of spin angular momentum measurements of electrons. The eigenvalues of the above generators are $\{ \frac{1}{2}, -\frac{1}{2}\}$. In the later chapters we will stick with $\sigma$ to represent the Pauli Matrices.
\\
\\
Finally, we check that $SU(2)$ and $SO(3)$ do have the same generator Algebra. Recall the definition of $X_{x}, X_{y}, X_{z}$ in the chapter 3.3, one can check that for any $i \neq j \neq k$
\begin{equation}
    [\sigma_{i}, \sigma_{j}] = 2i\epsilon_{ijk}\sigma_{k},
\end{equation}
following the same Lie Algebra.
\\
\\
Thus, the structral constants of the two Lie Algebra must be the same. We therefore conclude that the claim made at the end of last chapter about Algebra of $SU(2)$ and $SO(3)$ is the same, and the two dimensional representation of $SU(2)$ yields the two dimensional Lie Algebra.

\subsection{Isospin Symmetry}
Let's now focus on the two dimensional representation of $SU(2)$ and see what kind of physics can be this representation applies to. The first clear application is clearly the spin of the Fermions, given that the generators of $SU(2)$ are the Pauli Matrices. However, if spin angular momentum is the only symmetry we can talk about, this section wouldn't be called "Isospin Symmetry". So what indeed is isospin?
\\
\\
The story starts again from the great man of Heisenberg who claims, after the discovery of neutron in 1932 by the Nobel Prize Laureate James Chadwick \cite{Chadwick}, that the neutron and proton are the same particles in two different quantum states. The reason lies in the fact that though one of the particle is charged and the other one is uncharged, the rest mass of both particles look similar to each other. Thus, Heisenberg proposes that they should be considered invariant under the strong interaction, which is the interaction which creates forces to bind atomic nuclei. Thus the proton-neutron relation forms \textbf{doublets} symmetry, just as the spins of the electrons. Specifically, without any loss of generality, if we consider the \textbf{spinors} as the eigenstate of the Pauli Matrice $\sigma_{z}$, we get the following basis spinors to represent
\begin{align}
    p & = \begin{bmatrix}
        1 \\
        0
    \end{bmatrix} &
    n & = \begin{bmatrix}
        0 \\
        1
    \end{bmatrix}
\end{align}
\textbf{Definition 4.2.1} Multiplet. Multiplets are eigenstates of the generator of a symmetry. When there are two eigenstates, the multiplets are called doublets. And when there are three, they are called triplets etc.
\\
\\
\textbf{Definition 4.2.2} Spinors are analogy of Euclidean Vectors in Complex Space. In another word, they are just "vectors" under a transformation in Complex Space.
\\
\\
It is quite important to point out that, the action of $SU(2)$ on the spinors are different from the operation we perform previously for simulating a 3D unit rotation in $SU(2)$, as they are not 3D vectors but complex space spinors. Thus, for $t(\theta)$ in the set of $SU(2)$ using the same parametrization, we have
\begin{equation}
    v' = t(\theta) v = e^{i\theta^{\alpha} \sigma_{\alpha}} v 
\end{equation}
for some spinors $v$.
\\
\\
If we define 
\begin{equation}
    I_{+} = \frac{1}{2}(\sigma_{x} + i\sigma_{y}) = \begin{bmatrix}
        0 & 1 \\
        0 & 0
    \end{bmatrix},
\end{equation}
and
\begin{equation}
    I_{-} = \frac{1}{2}(\sigma_{x} - i\sigma_{y}) = \begin{bmatrix}
        0 & 0 \\
        1 & 0
    \end{bmatrix}.
\end{equation}
we can observe that
\begin{equation}
    I_{+}n = p,
\end{equation}
and
\begin{equation}
    I_{-}p = n.
\end{equation}
The operators $I_{+}, I_{-}$ are called the \textbf{ladder operators}, since $I_{+}$ raises $I_{+}$ the eigenstate of eigenvalue $-\frac{1}{2}$ to eigenstates of eigenvalue $+\frac{1}{2}$ and $I_{-}$ vice versa.
\\
\\
Physicists later figure out that the above proton-neutron symmetry belongs to some more fundamental symmetry of up and down quark favors. With simply the up quark being the state p and down quark being the state n. However, there are more quarks discovered in nature then the up and down quark, so the above symmetry is not enough to describe all symmetry involving quark favors. As a result, we need to move to higher dimensional symmetry such as $SU(3)$ (not higher dimensional representation of the same symmetry) to express these symmetries. 

\begin{figure}[h]
        \centering
        \includegraphics[height = 5.0cm]{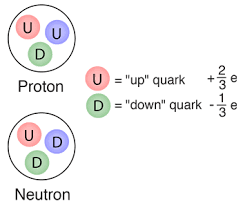}
        \caption{The neutrons and protons in terms of up and down quarks.}
        \label{pn}
    \end{figure}

\subsection{SU(3): Symmetry Breaking and Hypercharge}
Although isospin is proven to be indeed conserved in strong interactions of most hadronic particles, there are several particles such as the kaon whose strong interaction may not conserve isospins. Nobel Prize Laureate Murray Gell-Mann discover that these particles for which their strong interaction does not conserve isospin is also made up of another flavor of quark strangeness \cite{Gell}. Therefore, the $SU(2)$ symmetry which is used to express the symmetry of up and down quark flavors may not be useful anymore to describe the symmetry described here. Since the next higher-order symmetry is $SU(3)$ it is natural to think of to represent this new symmetry. 
\\
\\
In this case, we can use a smarter way to find the generators of $SU(3)$ than we use to find the generators of $SU(2)$, given we expect that the number of generators in $SU(3)$ can be much more than that in $SU(2)$. An important property of the generators of special unitary groups is that they have zero trace. That is the diagonal with these matrices sum to 0. We will provide a short proof of this fact in the following.
\\
\\
\textbf{Proof.} Let $U(t)$ be an arbitrary unitary operator with determinant 1, with a parameterization $U(0) = 1$. 
By the definition of generators $X_{\alpha}$, we know that
\begin{equation}
    U(t) = e^{iX_{\alpha}t}.
\end{equation}
Using the fact that for arbitrary operator $A(t)$
\begin{equation}
    det e^{A(t)} = e^{tr(A(t))}
\end{equation}
We find that
\begin{equation}
    1 = det \ U(t) = det \ e^{iX_{\alpha}t} = e^{tr(iX_{\alpha} t)}
\end{equation}
So
\begin{equation}
    it \cdot tr(X_{\alpha}) = 0, 
\end{equation}
and
\begin{equation}
    tr(X_{\alpha}) = 0.
\end{equation}
We then use this fact to claim that one parametrization of the generators are, given the generators are Hermitian:
\begin{equation}
    \begin{bmatrix}
        a & c - id & e - if \\
        c + id & b & g - ih \\
        e + if & g + ih & -a-b \\
    \end{bmatrix}.
\end{equation}
Thus, without loss of generality, we can write
\begin{equation}
    X_{\alpha} = \frac{1}{2} \lambda_{\alpha},
\end{equation}
where
\begin{align}
    \lambda_{1} & = 
    \begin{bmatrix}
        0 & 1 & 0 \\
        1 & 0 & 0 \\
        0 & 0 & 0 \\
    \end{bmatrix}, & \lambda_{2} & \begin{bmatrix}
        0 & -i & 0 \\
        i & 0 & 0 \\
        0 & 0 & 0 \\
    \end{bmatrix}, &
    \lambda_{3} & = \begin{bmatrix}
        1 & 0 & 0 \\
        0 & -1 & 0 \\
        0 & 0 & 0 \\
    \end{bmatrix}, \\ \lambda_{4} & = \begin{bmatrix}
        0 & 0 & 1 \\
        0 & 0 & 0 \\
        1 & 0 & 0 \\
    \end{bmatrix}, &
    \lambda_{5} & = \begin{bmatrix}
        0 & 0 & -i \\
        0 & 0 & 0 \\
        i & 0 & 0 \\
    \end{bmatrix}, & \lambda_{6} & = \begin{bmatrix}
        0 & 0 & 0 \\
        0 & 0 & 1 \\
        0 & 1 & 0 \\
    \end{bmatrix}, \\
    \lambda_{7} & = \begin{bmatrix}
        0 & 0 & 0 \\
        0 & 0 & -i \\
        0 & i & 0 \\
    \end{bmatrix}, & \lambda_{8} & = \frac{1}{3}\begin{bmatrix}
        1 & 0 & 0 \\
        0 & 1 & 0 \\
        0 & 0 & -2 \\
    \end{bmatrix},
\end{align}
where every $\lambda_{\alpha}$ is a unit determinant matrix. The $\lambda_{\alpha}$ are often time called the \textbf{Gell-Mann Matrices}. It can be observed that for $\alpha$ from 1 to 3
\begin{equation}
    \lambda_{\alpha} = \begin{bmatrix}
        \sigma_{\alpha} & \ & 0 \\
        \ & \ & 0\\
        0 & 0 & 0
    \end{bmatrix},
\end{equation}
for the $\sigma_{\alpha}$ being the pauli matrices, so we say that symmetry of isospin is contained in the symmetry of $SU(3)$ evident in the above construction. However, also note there is one weird lambdas, which is $\lambda_{8}$, which has eigenvalue $\{-\frac{2}{\sqrt{3}}, \frac{1}{\sqrt{3}}\}$, with degeneracy in $\frac{1}{\sqrt{3}}$, while the rest of the lambdas has eigenvalues $\{1, -1, 0\}$. Thus, for elements generated through $X_{8} = \frac{\lambda_{8}}{2}$ and some other $X_{\beta}$, we have must have a pair of eigenvalues instead of just one to represent a "quantum state". And this is where we can represent another conserved quantity besides isospin to compensate for the extra strange quark. Concretly, we call this conserved quantity \textbf{Hypercharge} denoted as Y, where under $SU(3)$ symmetry
\begin{equation}
    Y = B + S,
\end{equation}
for $S$ being the "strangeness" of a particle and $B$ being the baryon number. To see what are the possible quantum states, we do the same as for $SU(2)$ to define three basis spinors which are eigenstates of $\lambda_{3}$, the generalized version of $\sigma_{3}$.
\begin{align}
    e_{1} & = \begin{bmatrix}
        1 \\
        0 \\
        0
    \end{bmatrix}&  e_{2} &  = \begin{bmatrix}
        0 \\
        1 \\
        0
    \end{bmatrix} & e_{3} & = \begin{bmatrix}
        0 \\
        0 \\
        1
    \end{bmatrix},
\end{align}
Note that $X_{3}$ commutes with $X_{8}$ as both matrices are diagonal matrices, so the eigenstates of $X_{3}$ must also be eigenstates of $X_{8}$. Thus, we then apply $X_{3}$ and $X_{8}$ on the spinors and found that the three basis spinor represents states:
\begin{align}
    e_{1} = | \frac{1}{2}, \frac{\sqrt{3}}{6} \rangle \\ 
    e_{2} = | -\frac{1}{2}, \frac{\sqrt{3}}{6} \rangle \\ 
    e_{3} = | 0, -\frac{\sqrt{3}}{3} \rangle.
\end{align}
We can then use a \textbf{Weight Diagram} to visualize all pairs of eigenvalues as shown below. 

\begin{figure}[h]
        \centering
        \includegraphics[height = 5.0cm]{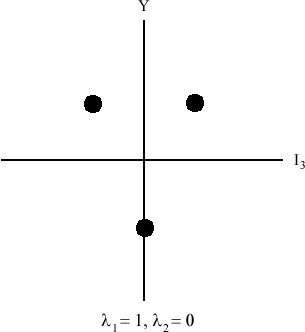}
        \caption{The weight diagram that shows the triplet of $SU(3)$ in its three-dimensional representation. Y is the hypercharge while $I_{3}$ is the z dimensional isospin quantum number, or the eigenvalues of $\sigma_{z}$}
        \label{pn}
    \end{figure}

\section{Lorentz and Poincare Groups}
So far we have been focusing on symmetry involving unit rotations, however, we know that there are also many other transformations such as translation and boost yet to be discussed. These transformations can be studied with classical mechanics under the usual Euclidean space structure but it would be better if we study them directly under the relativistic Minkowski spacetime framework, as classical transformations are a special limit to the relativistic transformations. Moreover, finding representations of the relativistic invariant, which we often call Lorentz or Poincare invariant, leads to operator algebra agreeing with relativity, and thus give us the freedom to investigate a relativistic theory of quantum mechanics, which we call \textbf{Quantum Field Theory}.

\subsection{Lorentz Group and its Four-Dimensional Representation}
\textbf{Definition 5.1.1} Lorentz Group is the symmetry of the Lorentz Transformations.
\\
\\
In the following, we will assume that the 0 rows of any four vector or four tensor as the time component, and 1, 2, 3 rows as the $x, y, z$ spatial components just as usual.
\\
\\
In special relativity, transformations are usually described under the axes time and the three spatial dimensions, so it is always the most natural to study the Lorentz Group with four-dimensional representation. In particular, the \textbf{Minkowski Spacetime Metric} $\eta$ specifies the geometrical structure of the spacetime.
\begin{equation}
    \eta = 
    \begin{bmatrix}
        -1 & 0 & 0 & 0 \\
        0 & 1 & 0 & 0 \\
        0 & 0 & 1 & 0 \\
        0 & 0 & 0 & 1 
    \end{bmatrix}.
\end{equation}
Thus, any inner product (dot product) of four vectors under the Minkowski spacetime must be written as
\begin{equation}
    x_{\mu}x^{\mu} = x^{\nu}\eta_{\nu\mu}x^{\mu}.
\end{equation}
The above inner product specifies a scalar which is invariant in any inertial reference frame, and we know that the Lorentz Transformation $\Lambda$ maps a vector from it measured in one inertial reference frame to another, we write
\begin{equation}
    x'^{\nu}\eta_{\nu\mu}x'^{\mu} = \Lambda^{\nu}_{\alpha} x^{\alpha}\eta_{\nu\mu} \Lambda^{\mu}_{\beta} x^{\beta} = x^{\alpha}\eta_{\alpha\beta}x^{\beta}.
\end{equation}
Canceling out the four vectors $x^{\alpha}$ and $x^{\beta}$, we obtain
\begin{equation}
    \label{solve}
    \Lambda^{\nu}_{\alpha} \eta_{\nu\mu} \Lambda^{\mu}_{\beta} = \eta_{\alpha\beta}.
\end{equation}
In matrix form, this writes
\begin{equation}
    \Lambda \eta \Lambda^{T} = \eta.
\end{equation}.
Now note that the determinant of $\eta$ is -1, and the determinant of a matrix's transpose equals the determinant of the matrix, we find that 
\begin{equation}
    -(det \Lambda )^{2} = -1, 
\end{equation}
so 
\begin{equation}
    det \Lambda = \pm 1.
\end{equation}
In addition, if we solve for Eq.\ref{solve} for the $\mu = \nu = 0$ component (exercise: check this), we find that
\begin{equation}
    \Lambda^{0}_{0} = \pm \sqrt{1 + \sum_{i} (\Lambda^{i}_{0})^{2}}.
\end{equation}
Notice that this implies
\begin{equation}
    |\Lambda^{0}_{0}| \geq 1.
\end{equation}
It is useful to split our Lorentz Transformations then into four categories
\begin{enumerate}
    \item $det \Lambda = 1$ and $\Lambda^{0}_{0} \geq 1$,
    \item $det \Lambda = -1$ and $\Lambda^{0}_{0} \geq 1$,
    \item $det \Lambda = 1$ and $\Lambda^{0}_{0} \leq -1$,
    \item $det \Lambda = -1$ and $\Lambda^{0}_{0} \leq -1$.
\end{enumerate}
The reason why it is useful is that one can observe only the first category can form a Lie Group given that all others does not contain the Identity Element. And if we call the Lorentz Transfromation Category which forms a Lie Group $\Lambda^{+}$, and define the Parity $T_{p}$ and Time Reversal Operators $T_{t}$ in 4D Minkowski Spacetime as
\begin{align}
    T_{p} &= \begin{bmatrix}
        1 & 0 & 0 & 0 \\
        0 & -1 & 0 & 0 \\
        0 & 0 & -1 & 0 \\
        0 & 0 & 0 & -1
    \end{bmatrix} & 
    T_{t} & = \begin{bmatrix}
        -1 & 0 & 0 & 0 \\
        0 & 1 & 0 & 0 \\
        0 & 0 & 1 & 0 \\
        0 & 0 & 0 & 1
    \end{bmatrix},
\end{align}
we find that the following set of Lorentz Transformations $\{ \Lambda^{+}, T_{p}\Lambda^{+}, T_{p}T_{t}\Lambda^{+}, T_{t}\Lambda^{+} \}$ along with the matrix multiplication operator does form a group. Moreover, one can check that the above four ways to write a Lorentz Transformation does describe all of the four categories of Lorentz Transformation we describe above. Thus, we say that the following set is the Lorentz Group, and the Lie Group formed by the $\Lambda^{+}$ is a \textbf{subgroup} of the Lorentz Group. By studying the generators of this subgroup we can then use $T_{p}$ and $T_{t}$ to find all other elements. 
\\
\\
Now we look at what are the elements of the subgroup formed by the $\Lambda^{+}$s. Let's start with examing the Lorentz Transformations with only spatial transformations, that is $\Lambda^{0}_{\nu} = \Lambda^{\mu}_{0} = 0$. Since for every element of $\Lambda^{+}$, we have $det \ \Lambda = 1$, and $\Lambda^{0}_{0} \geq 1$, the following three matrices can span all of the matrices of this kind, with the $\theta$ parametrization:
\begin{align}
    Rl_{x} & = \begin{bmatrix}
        1 & 0 & 0 & 0\\
        0 & 1 & 0 & 0\\
        0 & 0 & \cos\theta & -\sin\theta \\
        0 & 0 & \sin\theta & \cos\theta
    \end{bmatrix} & Rl_{y} &= \begin{bmatrix}
        1 & 0 & 0 & 0\\
        0 & \cos\theta & 0 & \sin\theta\\
        0 & 0 & 1 & 0 \\
        0 & -\sin\theta & 0 & \cos\theta
    \end{bmatrix}
    & Rl_{z} &= \begin{bmatrix}
        1 & 0 & 0 & 0\\
        0 & \cos\theta & -\sin\theta & 0\\
        0 & \sin\theta & \cos\theta & 0 \\
        0 & 0 & 0 & 1
    \end{bmatrix}
\end{align}
Note that this is exactly the four dimensional representation of $SO(3)$ rotating along the time axis, since the above three matrices are exactly
\begin{equation}
    Rl_{i} = \begin{bmatrix}
        1 & 0\\
        0 & R_{i}
    \end{bmatrix},
\end{equation}
for $R_{i}$ being the $SO(3)$ unit rotation along the i axis shown in chapter 3.3. The generators for $\Lambda$s in $\Lambda^{+}$ spanned by the Lorentz rotations are therefore
\begin{align}
    J_{x} &= \begin{bmatrix}
        0 & 0 & 0 & 0\\
        0 & 1 & 0 & 0\\
        0 & 0 & 0 & i \\
        0 & 0 & -i & 0
    \end{bmatrix} & J_{y} &= \begin{bmatrix}
        0 & 0 & 0 & 0\\
        0 & 0 & 0 & -i\\
        0 & 0 & 0 & 0 \\
        0 & i & 0 & 0
    \end{bmatrix} &
    J_{z} &= \begin{bmatrix}
        0 & 0 & 0 & 0\\
        0 & 0 & i & 0\\
        0 & -i & 0 & 0 \\
        0 & 0 & 0 & 0
    \end{bmatrix}
\end{align}
Given that they are generators for a four-dimensional representation of $SO(3)$, we called these generators the \textbf{Lorentz Rotation Generators}.
\\
\\
As you might have already thought of there are also elements of $\Lambda^{+}$ which can involve both spatial and time transformations. For these elements, it would be easier to derive their generators directly. Let us denote these generators by $K$ By the definition of generators, in some infinitesimal neighborhood of the identity element
\begin{equation}
    \Lambda \approx 1 - i\epsilon K,
\end{equation}
for some $\epsilon \approx 0$. We now plug in this definition of $\Lambda$ into Eq. \ref{solve}, 
\begin{align}
    (1 - i\epsilon K)\eta(1 - i\epsilon K)^{T} = \eta \\
    (\eta - i\epsilon K \eta)(1 - i\epsilon K^{T}) = \eta \\
    \eta - i\epsilon \eta K^{T} - i\epsilon K \eta - \epsilon^{2}K K^{T} = \eta.
\end{align}
By our assumption of $\epsilon \approx 0$, $\epsilon^{2} << \epsilon$, so
\begin{align}
    \label{someeq}
    - i\epsilon \eta K^{T} - i\epsilon K \eta = 0 \\
    \eta K^{T} = -K \eta.
\end{align}
Now we know that all Lorentz Transformations that involves both spatial and time transformations can be spanned by transformations involving x axis and time, transformations involving y axis and time,transformations involving z axis and time, just as we do for the unit rotations. We can then solve for each of the generators for these basis Lorentz Transformations. For example, for the generator $K_{x}$ generating the basis Lorentz Transformation involving x axis and time, every component except for $K_{x}^{0}{}_{1}, K_{x}^{1}{}_{1}, K_{x}^{1}{}_{0}, K_{x}^{0}{}_{0}$ of $K$ are identically 0. Using this fact and Eq. \ref{someeq}, we find that
\begin{equation}
    K_{x}^{1}{}_{0} = -K_{x}^{0}{}_{1} = i,
\end{equation}
for us want $K_{x}$ to have unit determinant and
\begin{equation}
    K_{x}^{1}{}_{1} = - K_{x}^{1}{}_{1} = K_{x}^{0}{}_{0} = K_{x}^{0}{}_{0} = 0.
\end{equation}
\begin{equation}
    K_{x} = \begin{bmatrix}
        0 & i & 0 & 0\\
        i & 0 & 0 & 0\\
        0 & 0 & 0 & 0\\\
        0 & 0 & 0 & 0
    \end{bmatrix}.
\end{equation}
Similarly, it will leave as an exercise for the readers to show that
\begin{align}
    K_{y} &= \begin{bmatrix}
        0 & 0 & i & 0\\
        0 & 0 & 0 & 0\\
        i & 0 & 0 & 0\\\
        0 & 0 & 0 & 0
    \end{bmatrix}, &
    K_{z} &= \begin{bmatrix}
        0 & 0 & 0 & i\\
        0 & 0 & 0 & 0\\
        0 & 0 & 0 & 0\\\
        i & 0 & 0 & 0
    \end{bmatrix}.
\end{align}
It is important to observe that the above generators are not Hermitian, so are not physical conserved charges. Given a theta parametrization, their corresponding basis Lorentz Transformations are
\begin{equation}
    B_{x}(\theta) = e^{iK_{x}\theta} = \begin{bmatrix}
        \cosh{\theta} & -\sinh{\theta} & 0 & 0\\
        -\sinh{\theta} & \cosh{\theta} & 0 & 0\\
        0 & 0 & 1 & 0\\
        0 & 0 & 0 & 1
    \end{bmatrix}
\end{equation}
\begin{equation}
    B_{y}(\theta) = e^{iK_{x}\theta} = \begin{bmatrix}
        \cosh{\theta} & 0 & 0 & -\sinh{\theta}\\
        0 & 1 & 0 & 0\\
        -\sinh{\theta} & 0 & \cosh{\theta} & 0\\
        0 & 0 & 0 & 1
    \end{bmatrix}
\end{equation}

\begin{equation}
    B_{x}(\theta) = e^{iK_{x}\theta} = \begin{bmatrix}
        \cosh{\theta} & 0 & 0 & -\sinh{\theta}\\
        0 & 1 & 0 & 0\\
        0 & 0 & 1 & 0\\
        -\sinh{\theta} & 0 & 0 & \cosh{\theta}
    \end{bmatrix},
\end{equation}
where $\sinh{\theta}$ and $\cosh{\theta}$ are the hyperbolic functions:
\begin{align}
    \sinh{\theta} & = \frac{e^{\theta} - e^{-\theta}}{2}, & cosh{\theta} & = \frac{e^{\theta} + e^{-\theta}}{2}.
\end{align}
The above transformations describe how the magnitude of a four velocity vector changes from one inertial reference frame to another. For example, consider the four velocity vector
\begin{equation}
    B_{x}(\frac{\pi}{2}) \begin{bmatrix}
        1 \\
        0 \\
        0 \\
        0 
    \end{bmatrix} = \begin{bmatrix}
        \cosh{\frac{\pi}{2}} \\
        -\sinh{\frac{\pi}{2}} \\
        0 \\
        0 
    \end{bmatrix},
\end{equation}
which means that an object originally at rest at a reference frame is now moving in another with speed
\begin{equation}
    \frac{dx}{dt} = \frac{dx}{d\tau}\frac{d\tau}{dt} = -\frac{\sinh{\frac{\pi}{2}}}{\cosh{\frac{\pi}{2}}}.
\end{equation}
Thus, we find that the operators spanned by the $B_{x}, B_{y}, B_{z}$ describes a Boost, or a change of speed of a four velocity vector, and we call the generators $K_{x}, K_{y}, K_{z}$ the \textbf{Lorentz Boost Generators}.
\\
\\
Finally, we take a look at what the Lie Algebra of the generators are. As an exercise, one can find the commutator for each pair of the above 6 generators and find that
\begin{equation}
    [J_{i}, J_{j}] = i\epsilon_{ijk}J_{k},
\end{equation}
\begin{equation}
    [J_{i}, K_{j}] = i\epsilon_{ijk}K_{k},
\end{equation}
\begin{equation}
    [K_{i}, K_{j}] = -i\epsilon_{ijk}J_{k},
\end{equation}
However, there's a more striking facts when we write 
\begin{equation}
    N^{\pm}_{i}  = J_{i} \pm iK_{i},
\end{equation}
we found that
\begin{equation}
    [N^{+}_{i}, N^{+}_{j}] = 2i\epsilon_{ijk} N^{+}_{k},
\end{equation}
\begin{equation}
    [N^{-}_{i}, N^{-}_{j}] = 2i\epsilon_{ijk} N^{-}_{k}.
\end{equation}
The above are exactly the Lie Algebra of $SU(2)$. In addition
\begin{equation}
    [N^{-}_{i}, N^{+}_{j}] = 0,
\end{equation}
making sure we do not leave the space. Thus, we find that the Lie Algebra of the subgroup formed by $\Lambda^{+}$ contains two copies of the Lie Algebra of $SU(2)$. 

\subsection{Weyl Spinors and Chirality}
We then move our attention to the two-dimensional representations of the Lorentz Group, or more simply its subgroup formed by $\Lambda^{+}$. Previously, we discover that the Lie Algebra of the $\Lambda^{+}$ subgroup contains two copies of the Lie Algebra of $SU(2)$ through expressing the Algebra of $N^{\pm}$, and we know that the two-dimensional representation of the $SU(2)$ generators are the Pauli Matrices $\sigma_{i}$, so it is natural to write the $Ns$ as the Pauli matrices. In the following, we will only consider the case when one of $N^{\pm}$ is the Pauli matrix as this is the case when we have more physical meanings.
\\
\\
Now let's start with assuming $N^{+}_{i} = \sigma_{i}$, and without loss of generality consider $N^{-}_{i} = 0$ satisfying the Lie Algebra of $\Lambda^{+}$ subgroup. Since
\begin{equation}
    N^{+}_{i} = J_{i} + iK_{i} = \sigma_{i},
\end{equation}
and
\begin{equation}
    N^{+}_{i} = J_{i} - iK_{i} = 0,
\end{equation}
we find that in this case our generators are
\begin{equation}
    J^{+}_{i} = \frac{\sigma_{i}}{2},
\end{equation}
and
\begin{equation}
    K^{+}_{i} = -i\frac{\sigma_{i}}{2}.
\end{equation}
We call the spinors acted by the generators and the Lorentz Group and Lorentz Boost operators in this kind the \textbf{Left-Chiral Spinors} denoted as $\chi_{L}$. 
\\
\\
Similarly, for the other case where $N^{+}_{i} = 0$ and $N^{-}_{i} = \sigma_{i}$, one can show that 
\begin{equation}
    \label{156}
    J^{-}_{i} = \frac{\sigma_{i}}{2},
\end{equation}
and
\begin{equation}
    \label{157}
    K^{-}_{i} = i\frac{\sigma_{i}}{2}.
\end{equation}
and all the spinors acted by the generators and the Lorentz Group and Lorentz Boost operators in this kind the \textbf{Right-Chiral Spinors} denoted as $\chi_{R}$. 
\\
\\
It is natural to ask why do we assign "left" and "right" to the naming of spinors. Now recall in the first chapter, when we define the "handedness" of the neutrinos using the Helicity of neutrinos, we also call the neutrinos "left-handed" or "right-handed", given they are mirror images under the parity operation. Thus, the "left" and "right" must also be related to some symmetry of the Parity Operator. Since we have defined the parity transformation $T_{p}$ in its four-dimensional representation, we can examine how parity acts on the generators in its four-dimensional representation and thus has the same effect in the two-dimensional representations. Since the parity operator must acts on the $\Lambda$s directly, one can check if we define
\begin{equation}
    J'_{i} = T_{p}J_{i}(T_{p})^{T},
\end{equation}
\begin{equation}
    \Lambda' = e^{iJ'_{i}\theta^{i}} = T_{p} \Lambda = T_{p} e^{iJ_{i}\theta^{i}}.
\end{equation}
and the same holds if we replace $J_{i}$ with $K_{i}$. Thus, the above action defines how the parity operator should act on the generators. Using the four-dimensional representations listed explicitly in Eq. 127, Eq. 136, and Eq. 137, we can further find that
\begin{equation}
    T_{p}J_{i}(T_{p})^{T} = J_{i},
\end{equation}
for every Lorentz unit rotation Generators, and
\begin{equation}
    T_{p}K_{i}(T_{p})^{T} = -K_{i}.
\end{equation}
That said, given Eq. \ref{156} and Eq. \ref{157}, where both $K_{i}$, $J_{i}$ are described of $\sigma_{z}$, if we are given a left-chiral spinor which is the eigenstate of both $K_{i}$, $J_{i}$
\begin{equation}
    \chi_{L} = |-\frac{1}{4}, \frac{i}{4}\rangle,
\end{equation}
the parity operation would take $\chi_{L}$ to some right-chiral spinor
\begin{equation}
    \chi_{R} = |-\frac{1}{4}, -\frac{i}{4}\rangle.
\end{equation}
We can observe that the "left and right" is exactly expressed by the eigenvalues of the $K_{i}$. We now have a new definition for the "handedness" using the eigenvalues of the Lorentz Group Generators in the above two dimensional representations. We call this new "handedness" the chirality to distinguish it from the helicity we discussed previously. The major difference between the two definitions of "handedness" is the following: for a massive particle moving with a speed less than the speed of light and is observed a positive helicity at frame S, and we can always find another reference frame moving faster than the particle measured at S where the helicity is negative, given that the helicity describes whether the direction of motion and the direction of the spin of a particle is the same; on the other hand, regardless of your reference frame, the chirality must be an invariant quantity as indicated by that the eigenvalues are conserved quantities of the Lorentz Group. Thus, helicity and chirality's notion of handedness only agrees with each other when we talk about massless particle which always moves at the speed of light. 
\\
\\
\textbf{Definition 5.2.1} Chirality is the pair of eigenvalues of the $(\frac{1}{2},0)$ and $(0, \frac{1}{2})$ representation of the Lorentz Group. And specifically, the eigenvalues of  $(\frac{1}{2},0)$ is the left chirality, and $(0, \frac{1}{2})$ is the right chirality.
\\
In the above definition, $(\frac{1}{2},0)$ representation specifies the representation where $N^{+} = \sigma$ and $N^{-} = 0$, and vice versa for $(0, \frac{1}{2})$.
\\
\\
\textbf{Definition 5.2.2} Weyl Spinors. Right-Chiral Spinors and Left-Chiral Spinors are called the Weyl Spinors.
\\
\\
The Weyl Spinors are important spinors in Quantum Field Theory as they described the spinors for the \textbf{Weyl Field Equation}, the wave equation for massless $\frac{1}{2}$ spin Fermions. Unfortunately, we will not discuss it here but always get to learn them when you study Quantum Field Theory in the future.

\subsection{Poincare Group and its Algebra}
If you are careful enough, you will see that there is one big problem about the Lorentz Group: not all operators in the group are Unitary Operators. In Quantum Mechanics, Unitary Operators are essentially required as the time evolution operator of a quantum state is unitary. Poincare Group is exactly the solution to this issue. It is proven that if we define Poincare Group as the group of Lorentz group "plus" the translation, we can find a finite unitary representation of this new group. Concretly, the translation group is
\begin{equation}
    T(v_{x}, v_{y}, v_{z}) = \begin{bmatrix}
        1 & 0 & 0 & v_{x}\\
        0 & 1 & 0 & v_{y}\\
        0 & 0 & 1 & v_{z}\\
        0 & 0 & 0 & 1
    \end{bmatrix}.
\end{equation}
where $v_{x}^{2} + v_{y}^{2} + v_{z}^{2} \leq c$. Note that "plus" isn't actually a numerical summation, it is a semidirect product of the translation group and the Lorentz Group. We will not discuss the semidirect product here but we encourage you to explore them and the rigorous mathematical definition of the Poincare Group in the future. However, this doesn't prevent us from presenting the Lie Algebra of the Poincare Group. Now Let's define $P_{i}$ as the generators of the translation group, we find out that the commutators of $P_{i}$ and each one of $K_{i}$ and $J_{i}$ is
\begin{equation}
    [J_{i}, P_{j}] = i\epsilon_{ijk}P_{k},
\end{equation}
and
\begin{equation}
    [J_{i}, K_{j}] = i\delta_{ij}P_{t},
\end{equation}
for $j$ being only spatial components $x, y, z$, where $\delta_{ij}$ is kronecker delta function
\begin{equation}
    \delta_{ij} = \begin{cases}
    0, & \mbox{if } i \ne j  \\
    1, & \mbox{if } i = j
    \end{cases}.
\end{equation}
And for the time component
\begin{equation}
    [J_{i}, P_{t}] = 0,
\end{equation}
and
\begin{equation}
    [K_{i}, P_{t}] = -iP_{i}.
\end{equation}
The commutation relation of $P$ itself is
\begin{equation}
    [P_{i}, P{j}] = 0.
\end{equation}
Of course this isn't a proof of all of the above commutation relations. The reason why we present them here is because of its prominent importance in Quantum Field Theory: in fact, Quantum Field Theory is the theory of fields invariant in unitary representation of Poincare Group. Thus, we believe this section is a good motivation for future study in Quantum Field Theory.

\newpage

\section{Noether's Theorem}
The final chapter would be devoted to one of the most important theorems discussing physical symmetry: Noether's Theorem. 
\\
\\
\textbf{Definition 6.1} Noether's Theorem (general version). For every continuous symmetry of the Lagrangian or Lagrangian density, there is a conserved current of that symmetry.
\\
\\
Noether's Theorem may sound more familiar to you if you have studied classical mechanics, and especially the Lagrangian Formalism of Mechanics, before. If you haven't studied Lagrangian mechanics, the textbook from John Taylor \cite{Taylor} would be a very good one to look it up. However, despite you have studied classical mechanics and work out the conserved current for a particular symmetrical transformation, it is very likely that the conserved current you work out belongs to the \textbf{particle theory} of Lagrangian Mechanics unless you have studied classical or quantum field theory before. In this chapter, we will focus our attention on the \textbf{field theory} version of Noether's Theorem and how it differs from the particle theory perspective. 

\subsection{Particle theory vs. Field theory}
In classical mechanics, we defined the Lagrangian of a system as
\begin{equation}
    L(q, \Dot{q}, t), 
\end{equation}
where $q$ is the generalized position of a physical object. What we want to solve for is the explicit form of $q(t)$, which is the equation of motion of the physical object. Such a formalism agrees with that of Newtonian Physics, which treat physical objects as approximately point masses and described the equation of motion by the three Newton's law. The theory that describes the physics of a particle in terms of its equation of motion is often called the particle theory of mechanics. This theory is the most helpful to describe physics in, for instance: classical object collisions, where the system's forces are mostly elastic forces and friction.
\\
\\
On the other hand, much physical interaction involves forces that manifest as field $\phi(\vec{x})$ distributed over the entire space. Note that we don't even have to talk about quantum mechanics to find such an example: the classical electromagnetic field and the gravitational field are examples of fields that are classical, though we believe quantum-mechanical counterpart of them might exist. In this case, it is best if our Lagrangian is a function of the entire field, which we often call \textbf{Lagrangian Density} 
\begin{equation}
    \mathcal{L}(\phi(\vec{x}), \partial_{\mu}(\phi(\vec{x})), \vec{x}, t),
\end{equation}
where $\partial_{mu}$ indicates the partial derivates are over each space and time coordinate. There are three major types of physical fields: scalar fields, vector fields, and tensor fields. An example of the scalar field is the Coulomb potential
\begin{equation}
    V(\vec{r}) = -\frac{kQq}{|r|},
\end{equation}
and its gradient, the electirc field is a vector field
\begin{equation}
    \vec{E}(\vec{r}) = \nabla V(\vec{r}).
\end{equation}
If we include relativity, we found that both electric field and the magnetic field describes the same force - the Lorentz force, so we can further group all the electric field and magnetic field together and form the electromagentic tensor field as following
\begin{equation}
    F^{\alpha \beta} = \begin{bmatrix}
0 & -E_x/c & -E_y/c & -E_z/c \\
E_x/c & 0 & -B_z & B_y \\
E_y/c & B_z & 0 & -B_x \\
E_z/c & -B_y & B_x & 0
\end{bmatrix}.
\end{equation}
When we work with the Lagrangian Density and Fields, we now observe that equation of motion is no longer relevant. It is the wave $\phi(x,t)$ for which matters propagate in the field that matters. Now we can bring back our quantum mechanics story: particles in quantum mechanics belong to wave theory but not the classical particle theory, and that's exactly the reason why studying everything in particle physics would be the best to study from the field theory, and is the point of our discussion here. 

\subsection{Noether's Theorem for Field Theory}
It is then a natural question to ask whether is also a "field version" of the general Noether's Theorem corresponding to the "particle version" with Lagrangian we saw in classical mechanics. In the following, though we won't be able to prove the theorem for an arbitrary symmetry, as it requires the study of \textbf{Gauge theory} for the internal symmetries which we didn't cover in this tutorial, we will show the conserved current of spacetime symmetries as a result of Noether's Theorem for you to grab a better understanding of the Theorem. However, it is still important for you to study internal symmetries if you would like to study Quantum Field Theory in the future, as an interacting field theory relies on internal symmetries, so I would recommend this book \cite{Cheng} to study Gauge theory.
\\
\\
Now let's start with a general Lagrangian Density $\mathcal{L}$ depends on some field $\phi$. A \textbf{spacetime tranformation} is some operator in a finite dimensional representation of the Poincare Group, where the Translation Group operation maps $\vec{x} \rightarrow \vec{x}'$ and the Lorentz Group operation maps $\phi \rightarrow \phi'$ through unit rotations or boost. Concretly, the spacetime transformation brings
\begin{equation}
    \phi(x_{\mu}) \rightarrow \phi(x_{\mu}) + \delta \phi,
\end{equation}
where we embed the time component and the spatial component into the same four covector $x_{\mu}$, and
\begin{equation}
    \delta \phi = \epsilon_{\mu\nu} S^{\mu\nu} \phi(x_{\mu}) - \partial_{x_{\mu}} \phi(x_{\mu}) \delta x_{\mu},
\end{equation}
where $\epsilon_{\mu\nu}$ is the Levi-Civita Symbol. The first part of the field variation $\delta \phi$ describes variation brought by the Lorentz Boost and unit rotation Generators, where $S^{\mu\nu}$ is related to $J_{i}$ by
\begin{equation}
    J_{i} = \frac{1}{2}\epsilon_{ijk}S_{jk},
\end{equation}
and $K_{i}$ is related to  $S^{\mu\nu}$
\begin{equation}
    K_{i} = S_{0i}.
\end{equation}
The second part is then brought by the translations.
\\
\\
The spacetime transformation maps the Lagrangian denisty:
\begin{equation}
    \mathcal{L}(\phi(x_{\mu}), \partial_{\mu} \phi(x_{\mu}), x_{\mu}) \rightarrow \mathcal{L}(\phi(x_{\mu}), \partial_{\mu} \phi(x_{\mu}), x_{\mu}) + \delta \mathcal{L},
\end{equation}
The variation of the Lagrangian density therefore writes:
\begin{align}
    &\delta \mathcal{L} = \frac{\partial \mathcal{L}}{\partial \phi} \delta \phi + \frac{\partial \mathcal{L}}{\partial(\partial_{\mu} \phi)} \partial_{\mu} \delta \phi + \frac{\partial \mathcal{L}}{\partial x_{\mu}} \delta x_{\mu} \\
    & = \partial_{\mu}(\frac{\partial \mathcal{L}}{\partial(\partial_{\mu} \phi)}\delta \phi) + \frac{\partial \mathcal{L}}{\partial x_{\mu}} \delta x_{\mu}.
\end{align}
Now let's consider when the variation of the field only contains translational variation, that is
\begin{equation}
    \delta \phi =  -\partial_{x_{\mu}} \phi(x_{\mu}) \delta x_{\mu}.
\end{equation}
where the $\delta x_{\mu}$ are simply some position four vectors in this case.Since a symmetry of the Lagrangian density means the variation $\delta \mathcal{L} = 0$,
we solve for
\begin{equation}
    \delta \mathcal{L} = -\partial_{\nu}(\frac{\partial \mathcal{L}}{\partial(\partial_{\nu} \phi)}\partial_{x_{\mu}} \phi(x_{\mu}) \delta x_{\mu}) + \frac{\partial \mathcal{L}}{\partial x_{\mu}} \delta x_{\mu} = -\partial_{\nu}(\frac{\partial \mathcal{L}}{\partial(\partial_{\nu} \phi)}\partial_{x_{\mu}} \phi(x_{\mu})  - \delta^{\nu}_{\mu} \mathcal{L}) \delta x^{\mu} =  0
\end{equation}
Since the position vectors $x^{\mu}$ are arbitrarily given, if we define the tensor
\begin{equation}
    T_{\mu}^{\nu} = \frac{\partial \mathcal{L}}{\partial(\partial_{\nu} \phi)}\partial_{x_{\mu}} \phi(x_{\mu})  - \delta^{\nu}_{\mu} \mathcal{L},
\end{equation}
we find that it must be the case
\begin{equation}
    \partial_{\nu} T_{\mu}^{\nu} = \partial_{t} T_{\mu}^{t} + \partial_{i} T_{\mu}^{i} = 0,
\end{equation}
where i are the spatial indexes. Finally, it follows directly from Eq. 186 that there are conserved quantities of the symmetry. First of all, since
\begin{equation}
    \partial_{t} T_{\mu}^{t} = - \partial_{i} T_{\mu}^{i} = -\nabla \vec{T},
\end{equation}
if we consider the quantity:
\begin{equation}
    Q_{\mu} = \int_{dV}  T^{0}_{\mu} d^{3}x,
\end{equation}
we find that, by \textbf{Divergence Theorem}
\begin{equation}
    \partial_{t} Q_{\mu} = \int_{V}  \partial_{t} T^{0}_{\mu} d^{3}x = \int_{V}   -\nabla \vec{T} d^{3}x = -\int_{\delta V}  \vec{T} d^{2}x
\end{equation}
Note that since the fields defining $T$ vanishes at $\infty$ to be physical fields
\begin{equation}
    \partial_{t} Q_{\mu} = -\int_{\delta V}  \vec{T} d^{2}x = 0.
\end{equation}
Thus $Q_{\mu}$ are the physical conserved quantities. Specifically, we define \textbf{total energy} as 
\begin{equation}
    E = Q_{0},
\end{equation}
and \textbf{linear momentum} as
\begin{equation}
    P_{i} = Q_{i}, \ \text{for i in 1 to 3.}
\end{equation}
We got the same conserved quantities as expected from the particle theory for translational symmetry. The tensor $T^{\nu}_{\mu}$ is therefore often called \textbf{Energy-Momentum Tensor} as it encodes information about the energy and momentum density and flux. 
\\
\\
The above derivation is just one example of conserved quantity for space-time symmetry when there is only translation. One can use the scalar field and the infinite dimensional representation of the Lorentz Group to show that
\begin{equation}
    L^{i}_{orbit} = \frac{1}{2} \epsilon^{ijk} \int (T^{j0} x^{k} - T^{k0} x^{j}),
\end{equation}
is the \textbf{total orbital angular momentum}, the conserved quantity for unit rotational invariance and 
\begin{equation}
    BO^{i} = \int (T^{00} x^{i} - T^{i0} x^{0}), 
\end{equation}
is the conserved quantity for the boost invariance. If you look for derivation, you can look at section 4.5.3 of \cite{Sch}. It would be best if you first understand the infinite-dimensional representation of the Lorentz Group to proceed.

\newpage

\section{Concluding Remarks and Acknowledgement}
Cheers! You have studied the mathematical prerequisite for understanding particle physics and are now ready to study Quantum Field Theory of QED, QCD, Electroweak theory! While this tutorial is single-handedly written by the author Jiaqi Huang myself, it has not been seriously peer-reviewed, so if you find out any issue in the tutorial feel free to contact me through email. 
\\
\\
For introductory-level textbook in particle physics and Quantum Field theory, I always recommend this book \cite{Griffiths} from Griffiths, one of the greatest physics educator, and this one \cite{Zee} for gentle introduction to Quantum Field Theory. If you want to study more about Poincare Group, you may take a look at \cite{Kim}. 
\\
\\
Finally, the tutorial is Jiaqi Huang's senior thesis at Carleton College. Many thanks to my particle physics Professor Dr. Chris West for his help in facilitating my understanding of the topic.

\newpage

%--BIBLIOGRAPHY--%

\end{document}